\documentclass[iop]{emulateapj}

\slugcomment{}

\shorttitle{A cosmic comb model of FRBs}
\shortauthors{Zhang}

\begin{document}


\title{FRB 121102: A repeatedly combed neutron star by a nearby low-luminosity accreting super-massive black hole}


\author{Bing Zhang}
\affil{Department of Physics and Astronomy, University of Nevada Las Vegas, NV 89154, USA}


%\altaffiltext{1}{Kavli Institute of Astronomy and Astrophysics, 
%Peking University, Beijing 100871, China}
%\altaffiltext{2}{Department of Astronomy, Peking University, 
%Beijing 100871, China}
%\altaffiltext{3}{Department of Physics and Astronomy, University of 
%Nevada Las Vegas, NV 89154, USA}


\begin{abstract}
The origin of fast radio bursts (FRBs) remains mysterious. Recently, the only repeating FRB source, FRB 121102, was reported to possess an extremely large and variable rotation measure (RM). The inferred magnetic field strength in the burst environment is comparable to that in the vicinity of the super-massive black hole Sagittarius A* of our Galaxy. Here we show that all the observational properties of FRB 121102 (including the high RM and its evolution, the high linear polarization degree, an invariant polarization angle across each burst and other properties previously known) can be interpreted within the ``cosmic comb'' model, which invokes a neutron star with typical spin and magnetic field parameters whose magnetosphere is repeatedly and marginally combed by a variable outflow from a nearby low-luminosity accreting super-massive black hole in the host galaxy. We propose three falsifiable predictions (periodic ``on/off'' states, and periodic/correlated variation of RM and polarization angle) of the model and discuss other FRBs within the context of the cosmic comb model as well as the challenges encountered by other repeating FRB models in light of the new observations.
\end{abstract}


\keywords{pulsars: general -- radiation mechanism: non-thermal -- radio continuum: general}



\section{Introduction}

Despite the rapid development in the observational front of fast radio bursts (FRBs) \citep{lorimer07,thornton13,petroff15,champion16,masui15,keane16,spitler16,delaunay16,chatterjee17}, we still do not know how these mysterious bursts are generated. Out of about two dozen FRB sources currently known, only one source, FRB 121102, was observed to repeat \citep{spitler16,scholz16,law17}, and it was precisely localized in a star-forming region within a low-metallicity dwarf galaxy at $z=0.193$ and is additionally associated with a persistent radio source \citep{chatterjee17,marcote17,tendulkar17,bassa17}. 

Recently, \cite{michilli18} reported some new observational results of FRB 121102 that brought important clues to understand the origin of this source.  These authors found that the radio emission of FRB 121102 is almost 100\% linearly polarized with an essentially constant polarization angle within each burst (but can vary among bursts). More intriguingly, these bursts have a very large value of Faraday rotation measure (RM) that varies in the range from $+1.46 \times 10^5$ to $+1.33 \times 10^5$ radians per square meter within seven months in the source reference frame. Such a large value of RM was discovered in the vicinity of the super-massive black hole in our galaxy, Sagittarius A*  and towards the active galactic nuclei (AGNs) in some galaxies \citep{bower03,marrone07}. \cite{michilli18} argued that the Faraday screen is local to FRB 121102, and estimated that the magnetic field strength along the line-of-sight is $B_\parallel = (0.6-2.4) f_{\rm DM}$, where $f_{\rm DM} > 1$ is a parameter to denote the ratio between the dispersion measure (DM) in the host and the DM that contributes to the RM. This magnetic field is orders of magnitude stronger than that in the interstellar medium, but is consistent with the environment in the vicinity of a super-massive black hole \citep{eatough13}. According to this picture, the steady radio source associated with FRB 121102 \citep{chatterjee17,marcote17} could be powered by a low-luminosity accreting super-massive black hole, and the surrounding star formation \citep{bassa17} could represent a circum-black-hole starburst \citep{michilli18}. 

Here we show that all the observations of FRB 121102 can be adequately interpreted within the framework of the ``cosmic comb'' model \citep{zhang17}. Within this model, an FRB is generated when an astrophysical gas flow (stream) interacts with the magnetosphere of a foreground 
neutron star. If the ram pressure of the stream exceeds the magnetic pressure at the light cylinder of the neutron star, the magnetosphere would be combed towards the opposite direction of the stream origin. As the combed magnetosphere sweeps the line of sight, an Earth-based observer detects an FRB. For FRB 121102, the source of the stream is the low-luminosity accreting super-massive black hole\footnote{In the original paper \citep{zhang17}, the source of the stream was not specified, even though a young magnetar was regarded as a plausible source.}, which sporadicly ejects a nearly isotropic disk wind outflow with a varying ram pressure during the accretion process \citep[e.g.][]{yuan12}.

%% In this section, we use  the \subsection command to set off
%% a subsection.  \footnote is used to insert a footnote to the text.


%% This section also includes several of the displayed math environments
%% mentioned in the Author Guide.

\section{The model}

\subsection{Model set-up}

\cite{michilli18} stated that the large RM value detected from FRB 121102 is similar to those seen towards massive black holes. For example, ${\rm RM} \approx -5 \times 10^5 \ {\rm rad \ m^2}$ is measured at $\sim 10^4$ Schwarzschild radii ($\sim 0.001$ pc) near the Milky Way's central black hole Sagittarius A*, and ${\rm RM} = -7 \times 10^4 \ {\rm rad \ m^2}$ is measured at a projected distance 0.1 pc ($\sim 10^6$ Schwarzschild radii) for the Galactic Center magnetar PSR J1745-2900  \citep{eatough13}. It is not known how magnetic field strength and configuration of super-massive black holes vary from case to case. Considering that the putative massive black hole is $\sim$ 2 orders of magnitude less massive than the Sagittarius A* black hole \citep{michilli18} and assuming that the magnetic field strengths of super-massive black holes are similar near the event horizon, one estimates that the physical distance of the neutron star from the putative black hole of FRB 121102 would be of the order of 0.001 pc. For easy discussion, in the following, we perform our quantitative estimations with the distance of the neutron star normalized to the fiducial distance $r_{_{\rm NS}} = 0.001$ pc from the central black hole\footnote{The discussion can be generalized to any distance based on the scaling laws with respect to $r_{_{\rm NS}}$.}. Consider a sporadic wind from the black hole with a typical dimensionless velocity $\beta = 0.01 \beta_{-2}$ (i.e. $3000 \ {\rm km \ s^{-1}}$), the ram pressure of the stream at $r_{_{\rm NS}}$ is \citep{zhang17}
\begin{equation}
P_{\rm ram} 
%= \rho v^2 = \frac{\dot M v}{4\pi r^2}
 \simeq  (160 \ {\rm erg \ cm^{-3}}) \left(\frac{\dot M}{\rm M_\odot \ yr^{-1}}\right) \beta_{-2} \left(\frac{r_{_{\rm NS}}}{10^{-3} \ {\rm pc}}\right)^{-2}.
\end{equation}
Requiring $P_{\rm ram} \gtrsim P_{\rm B,LC} = (B_s^2/8\pi)(\Omega R/c)^6$, one can constrain the neutron star parameters
\begin{equation}
B_{s,13}^2 P^{-6} \lesssim 46 \left(\frac{\dot M}{\rm M_\odot \ yr^{-1}}\right) \beta_{-2} \left(\frac{r_{_{\rm NS}}}{10^{-3} \ {\rm pc}}\right)^{-2}.
\end{equation}
Here $P$, $\Omega$, $B_s$, and $R$ are the period, angular frequency, surface magnetic field, and radius, respectively. Many Galactic pulsars satisfy such a condition. So the neutron star invoked in our model is a typical radio pulsar, which is otherwise undetectable in a distant galaxy.

\subsection{Data interpretation}

\begin{figure}
%\epsscale{.80}
\plotone{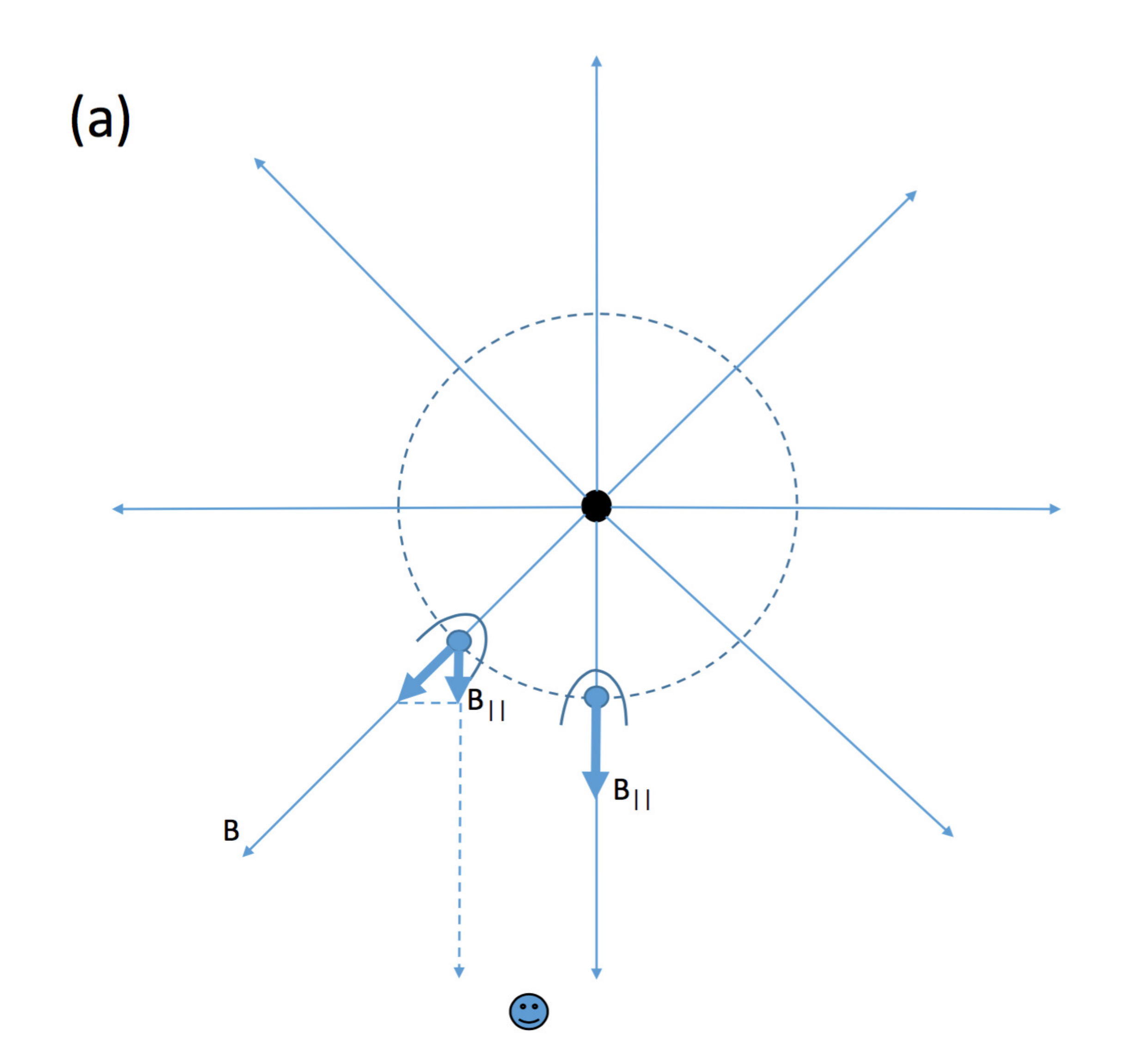}
\plotone{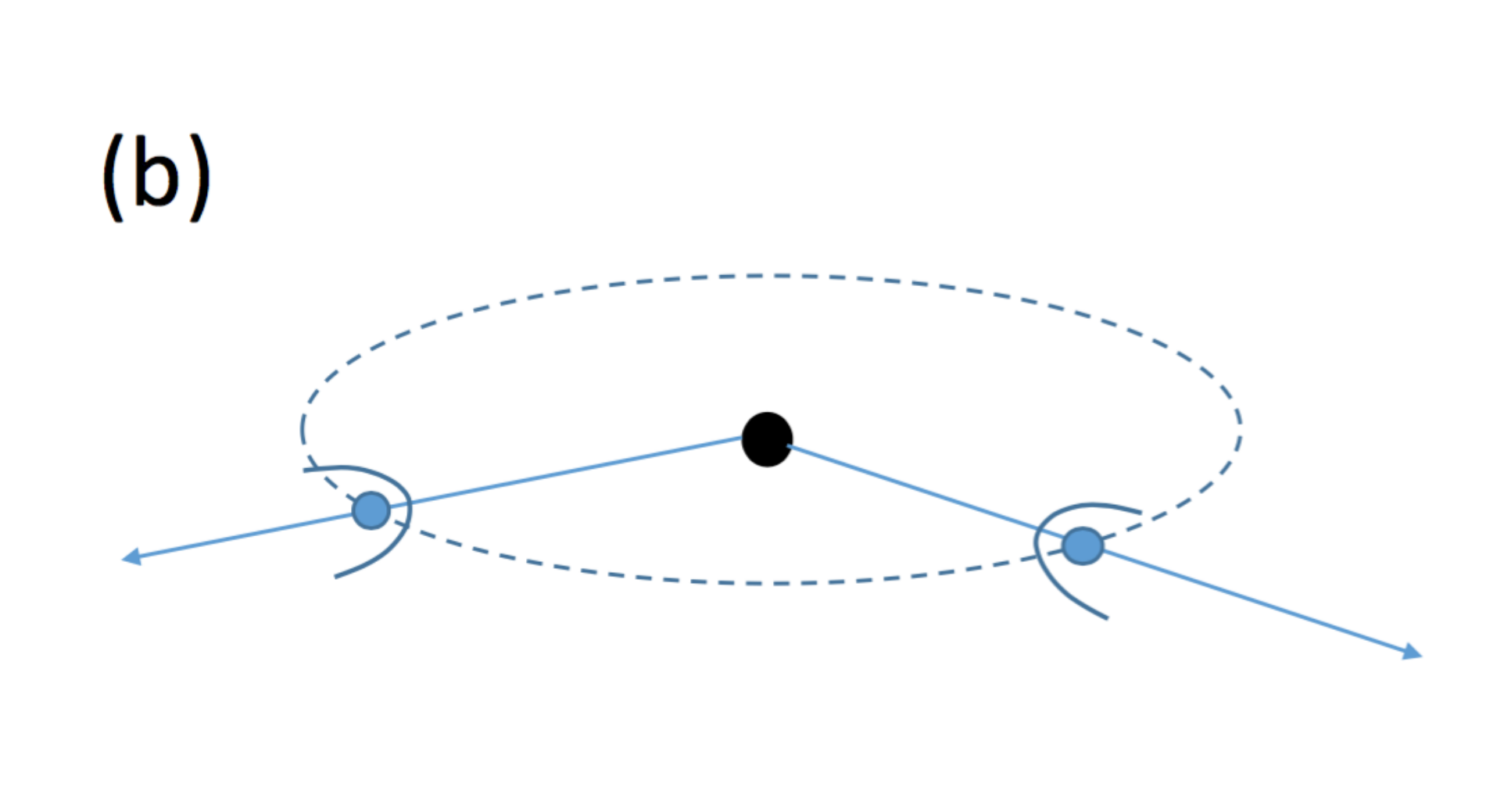}
\caption{A cartoon picture of the cosmic comb model for FRB 121102. (a) A face-on-orbit view: The outgoing arrows denote the projected magnetic field lines in the orbital plane. If the observer is off the plane, another $\cos i$ factor should be multiplied to obtain $B_\parallel$ measured by the observer. Two examples of the combing configurations are shown, which display different $B_\parallel$ component at the vicinity of the neutron star. As the neutron star move in the orbit around the super-massive black hole, a periodic variation of RM is expected. (b) An observer's view for two example combing configurations. The two arrows indicate the projected directions of the magnetic field lines when the combed beams sweep the direction of Earth. The polarization angle is constant for each combing event, but varies periodically as the neutron star moves in its orbit.}
\label{fig1}
\end{figure}

Such a set-up can account for all the observational data of FRB 121102:

\begin{itemize}
 \item {\em Large RM:} One may not be able to estimate the magnetic field strength near a super-massive black hole from first principles. However, in analogy with observations of the Galactic super-massive black hole Sagittarius A* \citep{bower03,marrone07,eatough13}, our set-up implies a magnetic field strength in the milli-Gauss range in the environment of the FRB source, which can account for the large RM as observed.
 \item {\em RM variation:} The variation of the RM value is about $(9-10)\%$ during a period of seven months. Within our model, this variation may be accounted for by the change of $B_\parallel$ integral due to the orbital motion of the neutron star around the black hole (Fig.\ref{fig1}a)\footnote{The magnetic field strength in the black hole vicinity is expected to decrease with radius rapidly (e.g. $B \propto r^{-3}$ for a dipolar configuration) so that the RM of the bursts are most sensitively related to the magnetic field strength and orientation at the immediate environment of the neutron star.}. For a black hole of mass $M_{\rm BH} \sim (10^4-10^6) M_\odot$ estimated based on a scaling relation between the black hole mass and the total stellar mass in the galaxy \citep{michilli18}, the orbital period of a neutron star at a distance $r_{_{\rm NS}}$ from the central black hole is
\begin{equation}
 P_{\rm orb} = 9.4 \ {\rm d} \left(\frac{r_{_{\rm NS}}}{10^{-3} \ {\rm pc}}\right)^{3/2} \left(\frac{M_{\rm BH}}{10^5 \ {\rm M_\odot}}\right)^{-1/2}.
\label{eq:Porb}
\end{equation}
In view of the uncertainty in $r_{_{\rm NS}}$ and $M_{\rm BH}$, there is a large parameter space where seven-month is of the order or much longer than $P_{\rm orb}$, so that significant RM variation is expected during the span of observations. Since the observations were not continuous, the Arecibo observations and the GBT observations likely picked up the neutron star at different orbital phases, and a $(9-10)\%$ variation of RM can be accounted for. Observationally, there is only a small ($\sim 0.2\%$) but systematic decrease of RM within the time scale of (1-2) days (when the first 15 bursts reported in Table 1 of \cite{michilli18} were discovered). On the other hand, a more significant decrease in RM is seen during the next two observational epochs spanning in the months time scale \citep{michilli18}. As a result, $P_{\rm orb}$ would be much longer than a day, but may not be much longer than the time scale of months. This is consistent with Eq.(\ref{eq:Porb}) given the uncertainty in both $r_{_{\rm NS}}$ and $M_{\rm BH}$. If the RM variation is mostly caused by the geometric effect (i.e. $B_\parallel$ integral variation as the neutron star orbits the black hole) rather than the fluctuation of the electron number density (which would also be associated with a variation in DM), then one would expect a periodic variation of RM in the time scale of weeks to months (period defined by $P_{\rm orb}/2$). Long-term monitoring of the source with RM measurements is encouraged to test such a prediction.
 \item {\em Linear polarization and non-varying polarization angle:} The emission mechanism of an FRB in the cosmic comb model is bunching coherent curvature radiation \citep{zhang17,yangzhang17b}.  The emission is expected to be highly polarized with the polarization angle defined by the direction of the magnetic field lines. Since in the combing model the magnetosphere of the neutron star is always combed from the black hole to the direction of the neutron star, the polarization angle is defined by the projection of that direction in the sky for each burst, which remains constant across the burst (Fig.\ref{fig1}b). Different bursts are produced as the neutron star is at different phases within the orbit, so that the polarization angle may vary from burst to burst\footnote{Due to the sporadic nature of the incoming streams that comb the neutron star, the evolution of the polarization angle may not be monotonic in a short period of times. However, in long term, one would observe a global trend of orbital evolution.}. For a nearly edge-on system, the polarization degree may vary moderately for most phases, but more significantly as the neutron star moves close to the line of sight. These are all consistent with the observations of FRB 121102 \citep{michilli18}. \cite{michilli18} disfavored the possibility that an emission beam sweeps across the line of sight based on the non-varying polarization angle. This is certainly a valid argument against the models invoking emission from the inner magnetosphere of a rotation-powered pulsar or magnetar \citep[e.g.][]{connor16,cordes16,metzger17,kumar17}. However, for the comb model this is not a concern, since the field line direction remains the same during each combing event, so that the duration of an FRB can be defined by the time when the combed beam sweeps across the line of sight \citep{zhang17}.
 \item {\em Repetition and temporal structure of the bursts:} FRB 121102 was observed to emit multiple bursts within the time span of several years. Within the cosmic comb model, the neutron star magnetosphere needs to be repeatedly combed. This requires that the outflow from the central black hole is unsteady with a variable velocity and density so that the ram pressure $P_{\rm ram} = \rho v^2$ fluctuates with time. As a stream with $P_{\rm ram} > P_{\rm B,LC}$ reaches the neutron star, the magnetosphere is combed to produce one burst. After the stream passes by, $P_{\rm ram}$ drops below $P_{\rm B,LC}$ and the magnetosphere would relax to the original configuration. Another burst is produced when another stream with $P_{\rm ram} > P_{\rm B,LC}$ arrives. The sporadic behavior of the bursts detected from FRB 121102 reflects the sporadic accretion behavior of the central black hole. Some repeating bursts from FRB 121102 have separations as short as $\Delta t \sim 20$ seconds. This requires that the spatial variation of the black hole outflow can be as small as $v (\Delta t) \sim 6 \times 10^{9} \beta_{-2} (\Delta t/20 \ {\rm s})$ cm. The time scale is shorter than the dynamical time scale at the black hole horizon, suggesting that the variability is caused by local small-scale processes in the disk wind, most likely due to magnetic reconnection \citep[e.g.][]{giannios09,zhang11}. Since the neutron star is repeatedly combed, given a certain range of $P_{\rm ram}$ variation, the combing events must be ``marginal'', i.e. $P_{\rm ram}$ is slightly greater than $P_{\rm B,LC}$ when a combing event happens. The magnetospheric structure of the neutron star is not completely removed. The produced FRB would have a temporal structure as an imprint of the original magnetospheric structure. This is consistent with the observed temporal features of the FRB 121102 bursts \citep{spitler16,michilli18}.
 \item {\em Non-varying DM:} Unlike a very young supernova remnant, the massive-black-hole-powered radio source is likely in a quasi-steady state within the time scale of years (e.g. in analogy to AGNs). Our model requires that the neutron star orbit (with a nominal radius of $\sim 0.001$ pc) is much smaller than the extent of the persistent radio source, the projected size of which is $\sim 0.7$ pc \citep{marcote17}. With such a configuration, the electron column density at the source likely remains essentially constant as the neutron star moves in its orbit. In principle, a small periodic variation of DM (with period $P_{\rm orb}/2$) is possible, but the amplitude of variation is much smaller than that of RM, so that it may not be detectable. 
  \item {\em Energy budget and luminosity of the bursts:} 
  The burst energy budget in the comb model ultimately comes from the accretion power of the super-massive black hole, which is essentially unlimited\footnote{This is different from the magnetar model whose energy budget is limited by the spin and magnetic energy of the neutron star.}. Within the theoretical framework of coherent curvature radiation by bunches, the luminosity (and brightness temperature) of an FRB depends on the fluctuating charge density in the magnetosphere (which scales with the local Goldreich-Julian density), the cross section, and the opening angle (of the order $1/\gamma_e$, where $\gamma_e$ is the Lorentz factor of electrons flowing inside the sheath) of the bunches. An advantage of the comb model is that the magnetic fields are combed to be nearly parallel to each other, so that the cross section of the bunch is much larger than the bunches from the polar cap region. The desired extremely high brightness temperature of FRBs is achievable with reasonable parameters without demanding a strong local magnetic fields (in contrast to the magnetar model). See Section 7.2 of \cite{yangzhang17b} for details.
 \item {\em Duration:} The duration of a burst is defined by the time scale when the combed emission beam sweeps the line of sight, which may be estimated as $ \Delta t \sim \frac{R_{\rm sh}} {v \gamma_e} \simeq (3.3 \ {\rm ms}) R_{\rm sh,9} \beta_{-2}^{-1} \gamma_{e,3}^{-1}$, where $R_{\rm sh}$ is the sheath radius, $\gamma_e$ is the typical electron Lorentz factor \citep{zhang17}.
 \item {\em Spectrum:} Given reasonable parameters, the typical frequency is in the GHz range \citep{zhang17,yangzhang17}. The spectral index in the high frequency regime varies from -1.3 to -3.3 for a reasonable value of electron energy spectral index. In the low-frequency regime, synchrotron self-absorption from the FRB-heated nebula would play a role to shape the spectrum and make a positive spectral index \citep{yang16,yangzhang17}. The predicted spectrum is therefore narrow. For different bursts, the peak frequency may vary slightly. This would result in a significant variation of the spectral indices in individual bursts, from steep positive spectral slopes (when the peak frequency is above the observational band) to steep negative spectral slopes (when the peak frequency is below the observational band), as observed in the bursts of FRB 121102 \citep[e.g.][]{spitler16,law17}. See Section 7.2 of \cite{yangzhang17b} for a more detailed discussion.
 \end{itemize}

\subsection{Falsifiable predictions}

This model has three falsifiable predictions that can be tested with future data:

\begin{itemize}
\item In order to have a combed beam sweep an Earth-based observer, Earth must be on the ``night'' side of the neutron star with respect to the super-massive black hole. As a result, only during half of the time in the neutron star orbit could repeating bursts be detected. The detected bursts should in principle have a $P_{\rm orb}/2$ period, but since another condition $P_{\rm ram} > P_{\rm B,LC}$ is needed to trigger a burst, one may not detect a periodic signal of the detected bursts due their sporadic nature. In any case, an ``on'' phase and an opposite ``off'' phase will alternate, even though it is possible to detect no bursts during the ``on'' phase. Applying an ``on-off'' template with different assumed $P_{\rm orb}$ to the available data may lead to a constraint on the allowed range of $P_{\rm orb}$. 
\item As explained above, this model predicts a periodic variation of the RM (with period $P_{\rm orb}/2$). For those bursts detected in the ``on'' phase, one could measure their RM and systematically search for possible periodicity of its variation to constrain $P_{\rm orb}$. Since the occurrence of the bursts is rather sporadic, very long term monitoring of the source is needed to verify this prediction.
\item Different bursts correspond to different phases in the neutron star orbit. One therefore predicts a periodic variation of the polarization angle with period $P_{\rm orb}/2$, even though it is constant in each individual burst (Fig.\ref{fig1}b). The orbital variations of polarization angle and RM should be correlated.
\end{itemize}

\section{Discussion}

We have shown that the currently available data of FRB 121102 can be adequately interpreted within the framework of the cosmic comb model \citep{zhang17}. In the following we discuss the implications of this conclusion for other FRBs and other FRB models.

\subsection{Other FRBs}

Thus far, FRB 121102 is the only FRB observed to repeat. One may speculate that other FRBs also repeat but their repeated bursts have not been detected. However, considering the non-detection limits of other FRBs and assuming that all FRBs are similar to FRB 121102, the probability that other bursts are not detected yet is found to be very low ($< 10^{-3}$), so that there could be more than one population of FRBs \citep{palaniswamy18}. Observationally, most non-repeating FRBs seem to have no temporal structure, with the width mainly defined by the scattering tail as the burst propagates in the interstellar/intergalactic medium \citep{keane16}.

It is possible that some non-repeating FRBs might be of a different physical origin, e.g. related to catastrophic events such as collapse of supra-massive neutron stars \citep{falcke14,zhang14} or mergers of compact objects \citep{totani13,zhang16a,wang16}. On the other hand, if all FRBs have the same physical origin, then those non-repeating FRBs may be understood in terms of strong (rather than marginal) combing events with $P_{\rm ram} \gg P_{\rm B,LC}$. Since the magnetosphere pressure is much smaller than the ram pressure, the imprint of the magnetosphere structure in the lightcurve would be diminished, so that the detected burst would not show a significant temporal structure. The magnetosphere hardly relaxes during the passage of the stream so that no repeating burst is detectable in short terms\footnote{The stream may also have a variable ram pressure. However, since $P_{\rm ram}$ is always much greater than $P_{\rm B,LC}$, no repeating bursts are expected.}. Another burst may be detected when another violent flare occurs. This would suggest a much longer waiting time than the typical waiting time of FRB 121102, consistent with the non-detection of repeating bursts despite intense searches \citep{petroff15b}. The astrophysical streams invoked in these events should be more violent. One example is FRB 150418 \citep{keane16}, whose bursting time coincided with an AGN flare in the field of view \citep{williams16,johnston17}. Since the chance probability of such an occurrence is quite low \citep{li16}, it is possible that FRB 150418 was actually produced by a foreground neutron star combed by the AGN flare \citep{zhang17}. The discovery of a possible super-massive black hole near FRB 121102 \citep{chatterjee17,marcote17,michilli18} greatly strengthened this possibility. 

Another example was a putative gamma-ray burst associated with FRB 131104 \citep{delaunay16,murase17,gaozhang17}. If the association is genuine, the FRB can be from a foreground neutron star combed by the blastwave of the GRB \citep{zhang17}.

\subsection{Other repeating FRB models}

The current data of FRB 121102 seem to pose great challenges to most other repeating FRB models discussed in the literature.

The leading model invokes a young magnetar that was born about a decade (or decades) ago, with the coherent radio emission powered by the spin energy or the magnetic energy of the magnetar \citep{katz16,chatterjee17,marcote17,tendulkar17,metzger17,kashiyama17}. The strongest support to the model was the resemblance of the FRB host galaxy with the host galaxies of long GRBs and superluminous supernovae \citep{tendulkar17,nicholl17}. However, with the high RM measurement and the possibility of a circum-black-hole starburst to interpret the data,  this initial motivation to invoke a young magnetar is no longer necessary. One may argue that the magnetar wind may provide the required $B_\parallel$ to interpret the large RM. However, such a high RM has never been observed in the vicinity of known magnetars unless it is close to the Galactic center \citep{michilli18}. Alternatively, one may invoke a young magnetar in the vicinity of the black hole and still require the magnetar itself to produce the bursts. However, the chance of having a young magnetar is much lower than having a typical pulsar near a super-massive black hole. One has to address the very small odds that the first young magnetar that generates repeating FRBs happens to be close to a super-massive black hole. In any case, in order to satisfy the energy and luminosity constraints from FRB 121102 using the magnetar energy budget (spin and magnetic energy), the magnetar cannot be too old. On the other hand, in order to allow GHz radio waves to escape freely and to avoid a detectable DM variation over a period of years, the magnetar cannot be too young. The young magnetar model is therefore subject to tight constraints in model parameters \citep{piro16,metzger17,cao17,kashiyama17,zhangzhang17,yangzhang17}. Interpreting the variation of RM and the constant polarization angle in each burst is also non-trivial, which requires an emission site near the light cylinder\footnote{The inter-pulses of the Crab pulsar have a flat polarization angle curve, and it is commonly suggested that the emission originates from an emission region close to the light cylinder \cite[e.g.][]{manchester05}.}. However, the extremely high brightness temperature of FRBs favors radio emission being produced in an emission region with strong magnetic fields close to the magnetar surface \citep{kumar17}. Such a model would predict a ``S'' or ``reverse-S'' shaped polarization angle evolution, and hence, is disfavored by the data.

Other models invoking an AGN to power FRBs \citep[e.g.][]{romero16,vieyro17} encounter great difficulties. In these models, FRBs are produced when a relativistic electron-positron beam hits ambient turbulent plasma clouds called cavitons to produce coherent radio emission through two-stream-instability-driven bunches. It is unclear whether such a coherent mechanism can produce the extremely high brightness temperature as observed in FRBs, and how a jet-cloud interaction may produce a narrow spectrum with a characteristic frequency in the GHz range. More severely, unlike curvature radiation in a pulsar magnetosphere, such emission is not expected to be polarized unless there is a local ordered magnetic field. Even if there is an ordered magnetic field in the medium, this field must be much weaker than that in a neutron star magnetosphere so that the emission must be greatly depolarized in the turbulent emission region. A near 100\% polarization degree of the bursts from FRB 121102 has ruled out such a scenario. The same argument also applies to other FRB models invoking a maser mechanism outside the magnetosphere of a neutron star \citep[e.g.][]{ghisellini17,waxman17,beloborodov17}.

Finally, the asteroid-neutron-star interaction model \citep{dai16,dai17,geng15} also needs to explain why such systems tend to stay close to a supermassive black hole, or why there is a large and variable RM. A high-RM model within such a scenario is being developed (Z.-G. Dai 2018, private communication).

In summary, the discovery of large and variable RM from FRB 121102 bursts \citep{michilli18} provides strong observational constraints to most repeating FRB models. As elaborated, the cosmic comb model can interpret all the available data so far and has three specific falsifiable predictions. Future long-term monitoring of the source as an effort of constraining the orbital period of the neutron star using the ``off'' phase and the periodicity of RM and polarization angle can eventually test this model.

\acknowledgments
I thank the referee for detailed comments and suggestions and Zi-Gao Dai, Derek Fox, 
Jason Hessels, Ye Li, Wenbin Lu, Rui Luo, Shriharsh Tendulkar, and Yuan-Pei Yang for helpful discussions.
This work is partially supported by NASA NNX15AK85G.

\end{document}